# Collision Resolution by Exploiting Symbol Misalignment

Lu Lu
Dept. of Information Engineering
The Chinese University of Hong Kong
ll007@ie.cuhk.edu.hk

Soung Chang Liew
Dept. of Information Engineering
The Chinese University of Hong Kong
soung@ie.cuhk.edu.hk

Shengli Zhang
Dept. of Information Engineering
The Chinese University of Hong Kong
slzhang5@ie.cuhk.edu.hk

*Abstract*—This paper presents CRESM, a novel collision resolution method for decoding collided packets in random-access wireless networks. In a collision, overlapping signals from several sources are received simultaneously at a receiver. CRESM exploits symbol misalignment among the overlapping signals to recover the individual packets. CRESM can be adopted in 802.11 networks without modification of the transmitter design; only a simple DSP technique is needed at the receiver to decode the overlapping signals. Our simulations indicate that CRESM has better BER performance than the simplistic Successive Interference Cancellation (SIC) technique that treats interference as noise, for almost all SNR regimes. The implication of CRESM for random-access networking is significant: in general, using CRESM to resolve collisions of up to $n$ packets, network throughput can be boosted by more than $n$ times if the transmitters are allowed to transmit more aggressively in the MAC protocol.

## I. INTRODUCTION

In wireless random-access networks, packet collisions are common. For example, in the popular IEEE 802.11 MAC, collisions occur when two or more stations decide to transmit to the access point (AP) simultaneously. At a station, a random backoff countdown process is used to decide when the station can transmit its packet. The most common cause of collisions is when two or more stations simultaneously count down to zero and transmit together. This can happen even when the stations can carrier-sense each other. Collisions can also happen due to the hidden-node phenomenon [1], wherein two stations that cannot carrier-sense each other transmit to the AP simultaneously.

This paper presents a novel method to recover collided packets. We call our method CRESM (collision resolution by exploiting symbol misalignment). CRESM does not require symbol-level synchronization among the stations. In fact, it thrives on symbol misalignment among the stations, which occurs naturally.

A fundamental concept underlying CRESM is that collided signals with symbol misalignment can be treated as the output from a virtual convolutional encoder. To the best of our knowledge, this is the first paper to use this concept to extract collided packets by means of (1) over-sampling and (2) an optimal Viterbi-like decoding algorithm.

*Related Work*

Ref. [2] proposes the disabling of the carrier sensing mechanism in a carrier-sense multiple access (CSMA) network to increase the likelihood of simultaneous transmissions (collisions). Collided signals are modeled using higher order constellation maps, and the joint decoding method requires symbol-level synchronization. Ref. [3] makes use of interference cancellation techniques to resolve the collisions. CRESM, on the other hand, does not assume symbol alignment. Also, CRESM does not require de-activating carrier sensing and can be deployed in a CSMA or a non-CSMA random access network.

In general, however, we do not advocate the disabling of carrier sensing when we can resolve collisions. Although we would want to encourage simultaneous transmissions, it is not clear that disabling carrier sensing *altogether* is the best way to do so. Instead, we would have better control over the system using a higher carrier sensing threshold or by increasing the transmission probabilities of the stations [4] in a way that is commensurate with the degree of collisions (number of packets in collisions) that can be dealt with.

Ref. [1] focuses on resolving collisions due to the hidden-node phenomenon. In Zig-Zag decoding [1], for example, several consecutive hidden-node collisions of the same group of packets are used to resolve the collided symbols. In practice, and in particular with the use of RTS/CTS, hidden-node collisions are not as common as backoff collisions. Furthermore, resolving hidden-node collisions does not boost the overall system throughput so much as it solves the unfairness problem induced by hidden nodes. Resolving backoff collisions, on the other hand, can potentially lead to much higher system throughput by allowing the stations to attempt to transmit more aggressively. CRESM can be used to deal with both backoff collisions and hidden-node collisions.

Recently, there have been intense research activities on using physical-layer network coding (PNC) [5, 6] to boost wireless network performance. The application domain of PNC is in relay networks. Here, we are interested in the more common WLAN scenario in which multiple stations want to transmit to a common access point (as the gateway to the Internet), and that the inter-traffic among the stations in the WLAN is minimal.

Finally, CRESM can be considered as a method for successive interference cancellation (SIC) [7, Ch.6] and multi-user detection (MUD) [8]. A distinguishing feature of CRESM is that it makes use of over-sampling on the unaligned overlapping packets to acquire more information on them.

## II. SYSTEM MODEL AND BASIC CRESM ALGORITHM

In this section we present the system model and the basic CRESM algorithm. To ease exposition, we describe CRESM in the terms of two-packet collisions and we assume carrier-phase synchronization between the two packets. Collisions of more than two packets and CRESM without carrier-phase synchronization will be treated in Section IV.

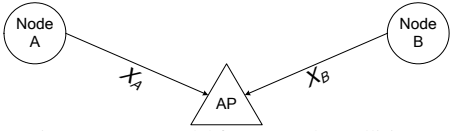

Fig. 1. System model for two packet collisions

### A. System Model

For a concrete picture, consider a CSMA wireless LAN as shown in Fig. 1. Suppose that both nodes *A* and *B* have packets to transmit to the AP. Nodes *A* and *B* first sense the channel to see if the channel is busy. Despite carrier sensing, it is still possible for *A* and *B* to transmit simultaneously when their backoff mechanism decides to transmit together. When that happens, the transmissions will collide at the AP.

We represent a wireless packet by a stream of discrete complex numbers. Specifically, we use complex numbers $x_A[m]$ and $x_B[m]$ to represent the modulated symbols of nodes *A* and *B* respectively. The overlapped signal received at the AP under packet collision for an AWGN channel is

$$y(t) = h_A(t)x_A[\lfloor t \rfloor]\cos(\omega_c t) \\ + h_B(t)x_B[\lfloor t-\Delta \rfloor]\cos(\omega_c(t-\Delta)) + w(t) \quad (1)$$

where $w(t)$ is Gaussian white noise with power spectral density $S_w(f) = N_0/2$; $h_A(t)$ and $h_B(t)$ are complex numbers that represent the channel attenuations with phase shift from *A* and *B* to the AP, respectively; $\lfloor t \rfloor$ is the largest integer no larger than *t*; $\omega_c$ is the carrier frequency; $\Delta$ is relative difference of the times of arrival of the two symbols at the AP. We assume that $h_A(t)$ and $h_B(t)$ stay constant throughout a packet duration. We further assume the transmit powers of nodes *A* and *B* have been combined into the corresponding $h_A(t)$ and $h_B(t)$.

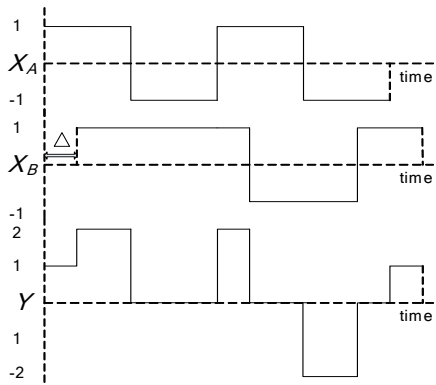

Fig. 2. Signals from *A* and *B* and the combined signal at the AP

In wireless CSMA protocols, there is typically no collision detection (e.g., 802.11). In the absence of collision detection, once the transmission of a packet begins, it will continue until the whole packet is transmitted, even while a collision is ongoing. An example of a collision of two BPSK modulated signals with perfect power control and carrier phase synchronization in continuous time is illustrated in Fig. 2.

### B. Discretization with Over-sampling

The basic structure of the receiver of CRESM is shown in Fig. 3. Over-sampling is used to generate two outputs in one symbol duration *T*. We assume normalization such that *T*=1 and $0 < \Delta < 1$ throughout this paper. The two output streams are then multiplexed into one discrete output stream $y[m]$, so that in this output stream there are two symbols per symbol duration, one from *A* and one from *B*.

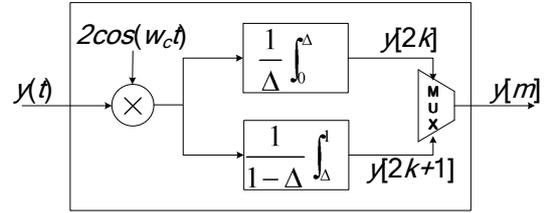

Fig. 3. Over-sampling at the receiver

For the over-sampling in CRESM, the integral of the traditional receiver [9], which integrates over the whole symbol duration, is modified and is now divided into two parts: one integral is from time 0 to time $\Delta$, and the other is from time $\Delta$ to time 1. The two discrete outputs (pre-MUX) of the receiver in Fig. 2 can be expressed as

$$y[2k] = \frac{2\int_0^\Delta y(t)\cos(\omega_c t)dt}{\Delta} = h_A x_A[k] + h_B x_B[k-1] + w[k]$$
$$y[2k+1] = \frac{2\int_\Delta^1 y(t)\cos(\omega_c t)dt}{1-\Delta} = h_A x_A[k] + h_B x_B[k] + w'[k] \quad (2)$$

for *k*=0,1,2,…, where $x_B[-1]=0$; and $w[k]$ and $w'[k]$ are Gaussian noises with variances of $N_0/2\Delta$ and $N_0/2(1-\Delta)$ respectively. In this and the following sub-sections, we assume perfect power control and carrier synchronization to ease exposition, i.e., $h_A=h_B=1$. Then the outputs in (2) can be simplified to (3)

$$y[2k] = x_A[k] + x_B[k-1] + w[k] \\ y[2k+1] = x_A[k] + x_B[k] + w'[k] \quad (3)$$

### C. Basic Idea of CRESM

First of all, we note that technically the two received packets will most likely arrive at the AP with symbol misalignment $\Delta$ if one does not deliberately try to synchronize the symbol arrival times. CRESM exploits this symbol misalignment to resolve collisions. Let us refer to the packets from *A* and *B* as $X_A$ and $X_B$, respectively. The symbols in $X_A$ are denoted by $x_A[0]x_A[1]x_A[2]...$, and the symbols in $X_B$ are denoted by $x_B[0]x_B[1]x_B[2]...$

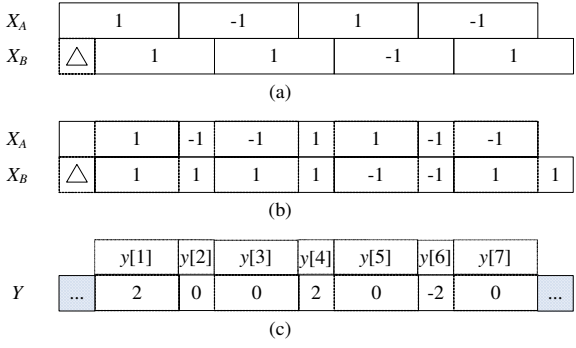

Fig. 4. The received signal at AP when there is no noise: (a) an example of specific symbol values; (b) over-sampled decomposition of overlapped symbols from *A* and *B*; (c) net superimposed symbol values received at AP.

The net effect of symbol misalignment is shown in Fig. 4, assuming the use of BPSK modulation (i.e., we map "0" bit to $e^{j0}=1$ and "1" bit to $e^{j\pi}=-1$). The effective "over-sampled" symbols as perceived at the AP are given in Fig. 4(c). CRESM makes use of these over-sampled symbols to recover the original symbols from *A* and *B*.

**Virtual Encoding**

Conceptually, CRESM treats $y[m]$ as the output from a "virtual" encoder $z[m]$ plus noise. Specifically,
$$z[2k] = x_A[k] + x_B[k-1]$$
$$z[2k+1] = x_A[k] + x_B[k] \quad (4)$$
where $x_A[k]$ and $x_B[k]$ are the $k^{th}$ symbols of $X_A$ and $X_B$ respectively. The possible values of the virtual encoder, i.e., the sum of the original symbols from *A* and *B*, are as follows:
$$1+1 = 2$$
$$1+(-1) = 0$$
$$(-1)+1 = 0 \quad (5)$$
$$(-1)+(-1) = -2$$
That is, two symbols in the domain of {1, -1} are encoded by the virtual encoder into one symbol with 3 possible values in the domain of {2, 0, -2}.

In the absence of noise, the possible values of the received sequence at the AP are the possible values of the virtual encoder's outputs as in (5). While a value of 2 (-2) in the output symbol means the original symbols from *A* and *B* are 1/1 (-1/-1), a value of 0 means the original symbols from *A* and *B* are either 1/-1 or -1/1 as in (6).
$$2 \to 1/1$$
$$0 \to 1/-1 \text{ or } -1/1 \quad (6)$$
$$-2 \to -1/-1$$
Thus, based on one virtual symbol alone, we cannot always recover the original symbols from *A* or *B*. However, with symbol misalignment and over-sampling receiver, an original symbol is actually mapped to two successive virtual symbols. Exploiting this "redundancy", the AP could recover the original symbols, with the following CRESM Algorithm.

**Successive Decoding**

CRESM employs a sort of successive decoding on the virtual symbols to recover the original symbols. The basic idea is illustrated in Fig. 5.

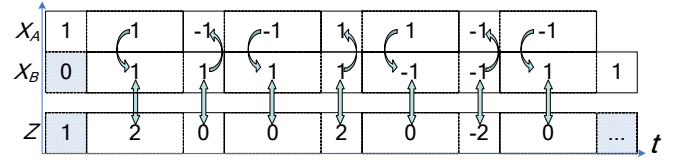

Fig. 5. The CRESM successive decoding

For the notation in (4), the decoding process can also be expressed as
$$x_A[k] = z[2k] - z[2k-1] + x_A[k-1]$$
$$x_B[k] = z[2k+1] - z[2k] + x_B[k-1] \quad (7)$$
where the current bit is actually decoded from the previous one with the knowledge of the overlapped symbols.

In the above paragraphs, we described the basic idea of CRESM in the absence of noise. In practice, noise is unavoidable and it can cause decoding error. To improve performance, we should make full use of the information received at the AP. For example, when the virtual symbol is 2 or -2, we do not need to rely on the previous virtual symbol to recover the original symbols as in (6). In Section III, we propose a sort of Viterbi-like decoding algorithm to decode the received signal to make full use of its information and to increase our confidence of correct detection.

### III. CRESM WITH VITERBI-LIKE DECODING

*A. Virtual Convolutional Encoding*

We can consider the virtual symbols as the output of a virtual convolutional encoder as shown in Fig. 6. The input to the virtual encoder consists of a stream of symbols which is a multiplexed stream of the original symbols from the two source nodes *A* and *B*. Specifically, the input stream is denoted by $v[0]v[1]v[2]v[3]v[4]v[5] = x_A[0]x_B[0]x_A[1]x_B[1]x_A[2]x_B[2]...$ where $x_A[0]x_A[1]x_A[2]...$ is the symbol stream from node *A* and $x_B[0]x_B[1]x_B[2]...$ is the symbol stream from node *B*. The virtual encoder in Fig. 6 produces an output stream $z[0]z[1]z[2]... = v[0](v[0]+v[1])(v[1]+v[2])...$. Obviously, they are the virtual symbols arrived at the receiver when there is no noise. To resolve the packet collision, the goal of our receiver is to recover the input stream of the virtual encoder, $x_A[0]x_B[0]x_A[1]x_B[1]x_A[2]x_B[2]...$, from which the individual streams from *A* and *B* can then be extracted.

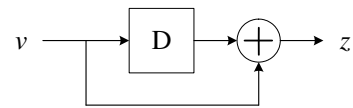

Fig. 6. The virtual convolutional encoder where D means one symbol delay

From Fig. 6 we notice that the virtual convolutional coding is the same as that of the conventional convolutional coding except for the following subtleties. First, the conventional convolutional code applies 'XOR' operation on the bits, but here we apply arithmetic addition. Second, in the conventional system, the output bits are transmitted with the same time duration and therefore the noise is statistically the same for each of the bits; in our system, the output bits have two possible durations: $\Delta$ for even bits, and $(1-\Delta)$ for odd bits, and consequently the noise levels for the bits alternate through the stream, due to the varying noise bandwidth during the sampling process. The effect of the alternating noise can be expressed mathematically as follows:

$$y[2k] = z[2k] + w[k]$$
$$y[2k+1] = z[2k+1] + w'[k] \quad (8)$$

where $w[k]$ and $w'[k]$ are the i.i.d Gaussian noises as introduced before, with variances of $N_0/2\Delta$ and $N_0/2(1-\Delta)$ respectively.

The virtual encoder in Fig. 6 corresponds to a convolutional code with code rate 1. That is to say there is no coding redundancy at all. The channel coding theory treats convolutional coding redundancy as a means for forward error correction (FEC). But since we introduce no coding redundancy here, CRESM itself cannot correct bit errors during transmission. However if the transmitted signals apply a FEC before modulation, the errors can be corrected after CRESM.

Treating the symbol in the register (Fig. 6) as having two possible states: '1' and '-1', the encoding process is given by Fig. 7, where $i/o$ denotes the input $i$ that triggers the transition, causing output $o$ to be produced. Let us assume that the register value is initialized to '1'. When the first symbol $v[0]$ arrives, the state could go from '1' to '-1' or stay at '1' depending on whether the this symbol is -1 or 1. If we further consider the possible transitions due to successive inputs over time, we get the virtual convolutional trellis as shown in Fig. 8, which is very similar to the conventional convolutional encoding process except that this virtual encoding is the conceptual outcome of simultaneously received symbols from two sources rather than FEC coding on bits. Although the motivation and the underlying phenomenon giving rise to CRESM are totally different from those of convolutional channel code, we can still apply a Viterbi-like decoding method to recover the originals bits in CRESM, as will be illustrated in the following subsection.

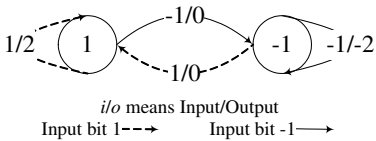

Fig. 7. The CRESM state transition diagram

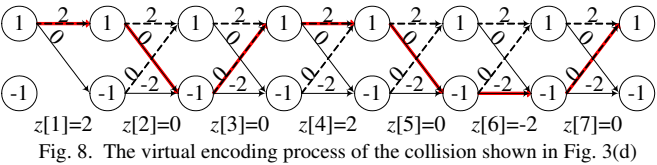

Fig. 8. The virtual encoding process of the collision shown in Fig. 3(d)

*B. Viterbi-like Decoding*

Maximum Likelihood (ML) decoding is optimal in terms of minimizing error probability when all input message sequences are equally likely. In particular, a ML decoder chooses $U^{(m^*)}$ if

$$P(Y | U^{(m^*)}) = \underset{\text{over all } U^{(m)}}{\arg\max} P(Y | U^{(m)}) \quad (9)$$

where $U^{(m)}$ is a possible input sequence and $P(Y | U^{(m)})$ is the *likelihood function* given the received sequence $Y$.

The Viterbi decoding algorithm, proposed and analyzed by Viterbi [10] in 1967, essentially performs ML decoding for convolutional code; however, it reduces the computational complexity by taking advantage of the special structure in the code trellis. Omura [11] demonstrated that the Viterbi algorithm is, in fact, an ML decoding method. The goal of selecting the optimal ML path can be expressed, equivalently, as choosing the codeword with the *minimum distance metric* [9, Ch. 7].

Inspired by Viterbi decoding, we propose a Viterbi-like decoding method for our virtual encoding process in Section IIIA. The details of our scheme and the original scheme are quite similar and can be expressed as follows. With reference to Fig. 7, once a symbol is received, we can calculate the distance from an originating state to a next state as the Euclidean distance $d_i$ between the received symbol $y_i$ and the corresponding transmitted symbol $z_i$ for that state transition. Different possible transitions correspond to different distances. We then store the cumulative distances $\sum d_i$ of different paths that correspond to different sequences of transitions. Given two paths with the same first and last states in their sequences, the path with the smaller accumulative distance is kept and the others are discarded. Thus, by computing the minimum distance path in the virtual coding trellis, we can then get $z[0]z[1]z[2]...$. After that, the original packets $X_A$ and $X_B$ can then be obtained using (7).

Since Viterbi decoding is a kind of optimal ML decoding method for convolutional code, we have the following proposition for our Viterbi-like decoding.

**Proposition 1**: Viterbi-like decoding is an ML decoding method among all possible CRESM decoding methods.

The proof of this theorem is similar to that of [11].

## IV. FURTHER DISCUSSION

In this section we discuss two extensions to CRESM. The first is the resolution of *n*-packet collision where *n* can be more than two. The second is CRESM without the assumption of carrier phase synchronization.

*A. CRESM with Collisions of More than Two Packets*

For collisions of more than two packets, the basic idea remains the same except that we need to increase over-sampling.

**Proposition 2**: The collision of *n* packets with symbol misalignment with $(n-1)$ different time shifts requires *n* samplings within one symbol.

For illustration, consider a 3-packet collision, say of nodes *A*, *B* and *C*. Let $X_A$, $X_B$ and $X_C$ be the corresponding packet-vectors containing information symbols. Then we have a virtual input streams $v[0]v[1]v[2]v[3]v[4]v[5]... = x_A[0]x_B[0]x_C[0]x_A[1]x_B[1]x_C[1]...$ to a virtual convolutional encoder as shown in Fig. 9. And the output of the virtual convolutional encoder is $z[-1]z[0]z[1]... = v[0](v[0]+v[1])(v[0]+v[1]+v[2])...$ which is used as the source for a Viterbi-like decoding method with 4 states. This virtual output stream Z can be obtained by three samples within one symbol.

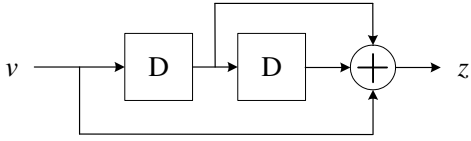

Fig. 9. The virtual encoder for CRESM with 3 collisions

The decoding process is similar to that of 2-packet collision except that the virtual convolutional encoding trellis has 4 states ('1,1'; '1,-1'; '-1,1'; and '-1,-1'). Due to the limited space here we omit the detailed description.

### B. CRESM without Carrier-Phase Synchronization

In this subsection we present a generalized version of CRESM (G-CRESM) that does not assume carrier-phase synchronization. We combine the effects of non-synchronization into the complex channel coefficients $h_A$ and $h_B$. The performance of G-CRESM is slightly better than CRESM with carrier-phase synchronization, as will be seen from our simulation results in Section V.

The coefficients $h_A$ and $h_B$ do not have any effect on the successive decoding algorithm in Section IIC. We can see this from the modified successive decoding outcomes

$$x_A[k] = \frac{1}{h_A}(z[2k] - z[2k-1] + x_A[k-1])$$
$$x_B[k] = \frac{1}{h_B}(z[2k+1] - z[2k] + x_B[k-1])$$
(10)

For Viterbi-like decoding in Section III, from (2) we can get

$$y_G[2k] = \frac{y[2k]}{h_A} = \frac{2}{\Delta \cdot h_A}\int_0^\Delta y(t)\cos(\omega_c t)dt$$
$$= x_A[k] + |\frac{h_B}{h_A}|e^{i(\varphi_B-\varphi_A)}x_B[k-1] + \frac{w[k]}{h_A}$$
$$y_G[2k+1] = \frac{y[2k+1]}{h_A} = \frac{2}{(1-\Delta)h_A}\int_\Delta^1 y(t)\cos(\omega_c t)dt$$
$$= x_A[k] + |\frac{h_B}{h_A}|e^{i(\varphi_B-\varphi_A)}x_B[k] + \frac{w'[k]}{h_A}$$
(11)

where $\varphi_B - \varphi_A$ is the relative phase of the $X_A$ and $X_B$.

We could rewrite the equations (11) as
$$y_G[2k] = x_A[k] + Hx_B[k-1]e^{i(\varphi_B-\varphi_A)} + n[k]$$
$$y_G[2k+1] = x_A[k] + Hx_B[k]e^{i(\varphi_B-\varphi_A)} + n'[k]$$
(12)

where $H = |h_B/h_A|$, and $n[k]$ and $n'[k]$ are the scaled Gaussian noises with variances of $\frac{N_0}{2\Delta|h_A|}$ and $\frac{N_0}{2(1-\Delta)|h_A|}$ respectively. The constellation map of G-CRESM in general has four points due to the phase difference of two transmissions, as illustrated in Fig. 10. The Euclidean distance of a received symbol is well defined based on this figure.

The decoding of G-CRESM is similar to that of CRESM described in Section IIIB; the only difference is the values of transition outputs in the virtual encoding trellis as shown in Fig. 11.

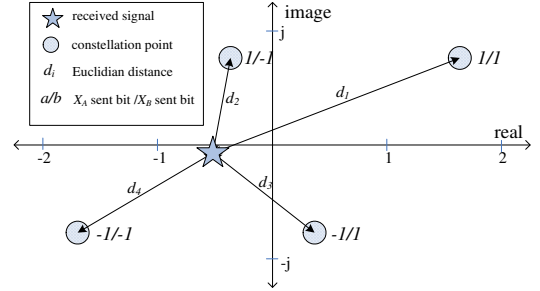

Fig. 10. Constellation map of G-CRESM with BPSK modulation and $\varphi_B - \varphi_A = \pi/4$

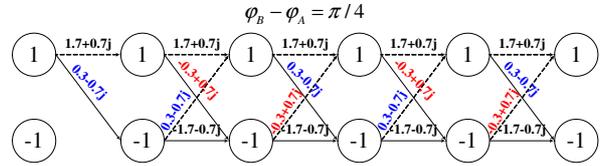

Fig. 11. Encoding trellis of G-CRESM

### V. SIMULATION RESULTS

We have performed simulations to investigate the performance of CRESM and G-CRESM. We assumed BPSK modulation with no channel coding. We used Viterbi-like decoding to resolve packet collisions, limiting our attention, however, to collisions involving two packets.

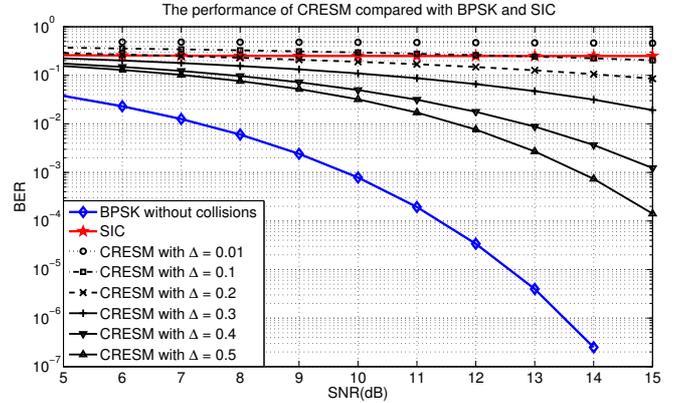

Fig. 12. The simulation results of CRESM with BPSK modulation

Fig. 12 shows the BER results. For benchmarking, we also present the results of non-simultaneous transmission with BPSK modulation, and SIC [3, 7, Ch.6] with the same transmission setup as in CRESM but which simply treats interference as noise during the decoding process. For CRESM, we can see that as $\Delta$ varies from 0 to 0.5 symbol length (and by symmetry, 1.0 to 0.5), the average BER decreases rapidly. For SIC, symbol misalignment has no effect because the power received from the other transmitter is simply treated as noise; furthermore, the BER does not improve with the increase of SNR since in our set-up the powers used by both transmitter are the same.

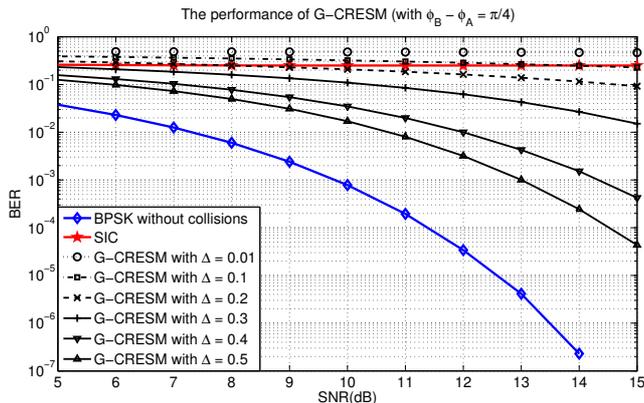

Fig. 13. The simulation results of G-CRESM with different $\Delta$

From Fig. 12 we also see that CRESM with half symbol misalignment ($\Delta = 0.5$) is 3.7 dB worse than the single-source BPSK case. This performance penalty from collision resolution is fundamentally due to two reasons. The first reason is over-sampling, which broadens the bandwidth of the noise by a factor of $\max\{1/\Delta, 1/(1-\Delta)\}$; as a result, the "effective" SNR is less than BPSK by at least 3dB. The second reason is the dependence of the decoding of successive bits; i.e., the decoding of the current bit depends not only on the current received symbol, but also on the previous decoded bit.

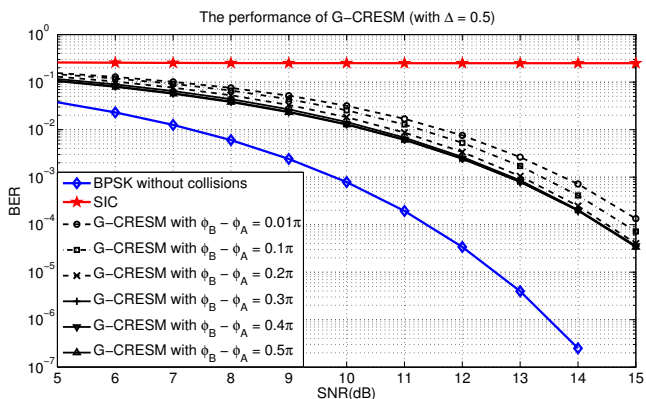

Fig. 14. The simulation results of G-CRESM with different relative phases

Fig. 13 shows the results of G-CRESM when the phases of the carriers of the two sources are not synchronized. We note that the BER results in Fig. 13 are actually better than those in Fig. 12, where the phases of the two sources are perfectly synchronized. Fig. 14 explores the phase affection on G-CRESM and the worst case is just the CRESM (G-CRESM with $\varphi_B - \varphi_A = 0°$). We notice that phase has smaller impact on the G-CRESM than $\Delta$. This means we do not need to deliberately synchronize the phase difference of the two sources: i.e., using G-CRESM we could deal with the phase asynchrony at the receiver.

## VI. CONCLUSION AND FUTURE WORK

This paper has proposed and investigated CRESM, a novel packet collision resolution scheme that treats collided signals as of the output of a virtual convolutional encoder. The main essence of CRESM is that by over-sampling the overlapping signals at the receiver, one could extract the individual packets from the transmitters. Within this general construct, we propose a specific Viterbi-like decoding scheme that minimizes the BER. As far as we know, this is the first paper that proposes to treat collisions as a kind of convolutional code, to which simple digital signal processing (DSP) techniques could then be applied for decoding purposes.

Although we have described CRESM in the context of 802.11, the idea behind CRESM is in fact quite general and is applicable to other MAC protocols (e.g., Aloha). Within a larger context, CRESM can be viewed as a technique for multiple-packet reception (MPR) [12]. An attractive feature of CRESM is that no symbol-level synchronization is required of the simultaneously transmitting stations – in fact CRESM exploits the naturally occurring symbol misalignment to perform MPR.

Given that collisions can be resolved in a simple manner at the physical layer by CRESM, an implication is that the MAC protocol should be redesigned in such a way as to encourage multiple packet transmissions. That is, the stations should be more aggressive in their transmissions. For example, in [4], it was shown that the network throughput can be increased by $n$ times if one could resolve $n$-packet collisions.

In this paper we have only investigated the collisions of BPSK packets. The idea of CRESM, however, is independent of modulation. As extension work, it will be interesting to investigate CRESM using an information-theoretic approach.

REFERENCES

[1] S. Gollakota and D. Kattabi, "ZigZag Decoding: Combating Hidden Terminals in Wireless Networks," *MIT-CSAIL-TR-2008-018* April 8, 2008.
[2] D. Halperin, T. Anderson, and D. Wetherall, "Interference Cancellation: Better Receivers for a New Wireless MAC," *Hotnets* 2007.
[3] D. Halperin, T. Anderson, and D. Wetherall, "Taking the String out of Carrier Sense: Interference Cancellation for Wireless LANs," *ACM MOBICOM* 2008, San Francisco, USA.
[4] P. X. Zheng, Y. J. Zhang, and S. C. Liew, "Analysis of Exponential Backoff with Multipacket Reception in Wireless Networks," The 6th *IEEE Workshop on Wireless Local Networks*, Nov 2006.
[5] S. Zhang, S. C. Liew, and P. K. Lam, "Physical Layer Network Coding," *ACM MOBICOM* 2006, Los Angeles, USA.
[6] S. Katti, S. Gollakota, and D. Kattabi, "Embracing Wireless Interference: Analog Network Coding," *ACM SIGCOMM* 2007, Kyoto, Japan.
[7] D. Tse and P. Viswanath, *Fundamentals of Wireless Communication*, Cambridge University Press, 2005
[8] S. Verdu, *Multiuser Detection*, Cambridge University Press, 1998
[9] B. Sklar, *Digital Communications: Fundamentals and Applications* (second editon) Prentice-Hall PTR, 2001.
[10] A. J. Viterbi, "Error Bounds for Convolutional Codes and an Asymptotically Optimum Decoding Algorithm," *IEEE Transactions on Information Theory*, vol. IT-13, April, 1967, pp. 260-269.
[11] Omura, J. K., "On the Viterbi Decoding Algorithm" *IEEE Transactions on Information Theory*, vol. IT-15, Jan. 1969, pp. 177-1769.
[12] P. X. Zheng, Y. J. Zhang, and S. C. Liew. "Multipacket Reception in Wireless Local Area Networks," *IEEE ICC* 2006, Istanbul, Turkey